\documentclass[aps,amssymb,amsmath,twocolumn, pra, superscriptaddress]{revtex4}
\usepackage{graphicx}
\usepackage{epsfig}
\usepackage{dcolumn}
\usepackage{bm}
\usepackage{bbm}
\usepackage{hyperref}
\usepackage[up]{subfigure}

\newcommand{\be}{\begin{equation}}
\newcommand{\ee}{\end{equation}}
\newcommand{\bc}{\begin{center}}
\newcommand{\ec}{\end{center}}
\newcommand{\bi}{\begin{itemize}}
\newcommand{\ei}{\end{itemize}}
\newcommand{\bea}{\begin{eqnarray}}
\newcommand{\eea}{\end{eqnarray}}
\begin{document}
\title{Quantum phase transition using quantum walks in an optical lattice}
\author{C. M. \surname{Chandrashekar}} 
\affiliation{Institute for Quantum Computing, University of Waterloo, ON, N2L 3G1, Canada}
\author{Raymond \surname{Laflamme}}
\affiliation{Institute for Quantum Computing, University of Waterloo, ON, N2L 3G1, Canada}
\affiliation{Perimeter Institute for Theoretical Physics, Waterloo, ON, N2J 2W9, Canada}
\begin{abstract}
We present an approach using quantum walks (QWs) to redistribute ultracold atoms in an optical lattice. Different density profiles of atoms can be obtained by exploiting the controllable properties of QWs, such as the variance and the probability distribution in position space using quantum coin parameters and engineered noise. The QW evolves the density profile of atoms in a superposition of position space resulting in a quadratic speedup of the process of quantum phase transition. We also discuss implementation in presently available setups of ultracold atoms in optical lattices.  
\end{abstract}
\maketitle
\preprint{Version}
\section{Introduction}

Exploiting the aspects of quantum mechanics, such as superposition 
and interference, has lead to the idea of quantum walks (QWs), a generalization of classical random walks (CRWs)  \cite{riazanov, feynman, aharonov, meyer96, meyer2}. In the CRW the particle moves in the configuration space with a certain probability, whereas  
in the QW the particle moves in a superposition of the 
configuration space with a probability amplitude. Because of the quantum interference effect, the 
variance in position $\sigma^{2}$ of the QW is known to grow quadratically with the 
number of steps $N$, $\sigma^{2}\propto N^{2}$, compared to the linear growth, 
$\sigma^{2}\propto N$, for the CRW. This motivated the research in the direction of finding 
quantum algorithms with optimal efficiency using the  QW~\cite{childs, shenvi, childs1, ambainis}.
Experimental implementation of the QW has also been reported \cite{ryan, du, perets}.
\par
The  continuous time quantum walk (CTQW) and the discrete time quantum 
walk (DTQW) are the two widely  studied versions of the QW \cite{kempe}.   
In the CTQW \cite{farhi}, one  can directly define the walk on the position  
space,  whereas,  in  the DTQW  \cite{andris}, it is necessary to introduce 
a quantum coin operation to define the direction in which the particle  
has to move. Due to the coin degree of freedom, the discrete time variant is shown to be more  
powerful than the other  in  some contexts  \cite{ambainis}, and  the coin parameters (scattering operator) 
can be  varied to  control the dynamics of the evolution \cite{meyer96, chandra08}. Therefore, in this paper we will consider only the discrete time variant of the QW.  Beyond quantum computation, the QW can be  used to demonstrate coherent quantum control over, for example, atoms, photons or spin chain 
systems. Along with purely quantum dynamics, a small amount of engineered noise or 
environmental effect can enhance the properties of the QW \cite{kendon, chandra07, brun}. 
As we show later this can enlarge the toolbox for controlling the dynamics in a physical system. 
\par
This paper describes the use of a QW which consists of an iteration of the quantum coin operation 
and a subsequent conditional shift operation to control and study the dynamics of ultracold bosonic atoms in an optical lattice. The quantum coin operation evolves the atom wave function into the superposition of the internal state of the particle and the shift operation spreads the atom  wave function in 
a superposition of the position space. In particular, we show that the quantum phase transition from the Mott insulator (MI) - a regime where no phase coherence is prevalent - to the superfluid (SF) - a regime with 
long-range phase coherence \cite{jaksch, greiner} and vice versa can be obtained by redistributing the density profile of atoms using a QW. The simulation of the quantum phase transition using the QW occurs quadratically faster in one dimension (1D) compared to, varying the optical lattice depth and letting the atom-atom interaction follow the CRW behaviour. Enhancing the properties of the QW by varying the quantum coin parameters or by addition of experimentally engineered noise in small amounts \cite{chandra07} can be used as an additional tool to evolve the atoms into different number density distributions. 
\par
Theoretical studies of the dynamics of atoms in an optical lattice are done using mean field approaches \cite{van} and quantum Monte Carlo methods \cite{batrouni}. The quantum  correlation induced between atoms and position space by the QW can serve as an alternate method for theoretical studies. The use of the QW to study the dynamics of particles in 1D magnetic systems \cite{osterloh}, and photonic Mott insulators \cite {hartmann} or to observe complex quantum phase transitions \cite{demler} and quantum annealing \cite{somma} would be of both theoretical and experimental relevance.   
\par
This paper is organized as follows. In Sec. \ref{dtqw} we define the DTQW and its properties, and in Sec. \ref{pt} the phase transition in an optical lattice is discussed. The implementation of the QW on atoms in 1D MI and SF regimes is discussed in Sec. \ref{impQW},  where we analyze the dynamics and present the density profile obtained by exploiting some of the properties of the QW.  In Sec. \ref{qwnoise} we discuss the 
QW with a noisy channel as a tool to control the redistribution of atoms. In Sec. \ref{impl} we briefly discuss 
a scheme for experimental realization before concluding in Sec. \ref{conc}.  

\section{Discrete time Quantum Walk}
\label{dtqw}

To define a 1D DTQW we require the {\it coin} Hilbert space $\mathcal H_{c}$ and the {\it position} Hilbert space $\mathcal H_{p}$. $\mathcal H_{c}$ is spanned by the internal state of the particle, $|0\rangle$ and $|1\rangle$, and $\mathcal H_{p}$ is spanned by the basis states $|j\rangle$, $j \in \mathbb{Z}$. The total system is then in the space $\mathcal H = \mathcal H_{c} \otimes \mathcal H_{p}$. To implement the QW, the particle at the origin ($j=0$) in state $|j_{0}\rangle$ is evolved into a superposition of $\mathcal H_{c}$, 
\be
|\Psi_{\mbox{in}}\rangle=\frac{1}{\sqrt 2}[|0\rangle+i|1\rangle]\otimes|j_{0}\rangle
\ee
and then subjected to the conditional shift operation $S$ to evolve into a superposition in the position space:
\begin{eqnarray}
\label{eq:condshift}
S = |0\rangle \langle 0|\otimes \sum_{j \in \mathbb{Z}}|j-1\rangle \langle j |+|1\rangle \langle 1 |\otimes \sum_{j \in \mathbb{Z}} |j+1\rangle \langle j| \nonumber \\
\equiv  |0\rangle\langle0|\otimes \hat{a} + |1\rangle\langle1|\otimes \hat{a}^{\dag}.
\end{eqnarray}
Here $\hat{a}$ and $\hat{a}^{\dag}$ are annihilation and creation operations acting on $\mathcal H_{p}$. 
The $S$ is followed by the quantum coin operation, which in general can be written as an $SU(2)$ operator of the form
\be
B_{\xi,\theta,\zeta} =\left( \begin{array}{clcr}
 e^{i\xi}\cos(\theta)  & &   e^{i\zeta}\sin(\theta)   \\
e^{-i\zeta} \sin(\theta)  & &  -e^{-i\xi}\cos(\theta)
\end{array} \right),
\ee
to evolve the particle into a superposition in $\mathcal H_{c}$. Therefore, each step of the QW is composed of an application of operation $B$ and a subsequent operation $S$ to entangle $\mathcal H_{c}$ and $\mathcal H_{p}$. The process of application of $W=S(B_{\xi,\theta,\zeta} \otimes \mathbbm{1})$ is iterated without resorting to intermediate measurements to realize large number of steps. 
\begin{figure}[h]
\begin{center}
\epsfig{figure=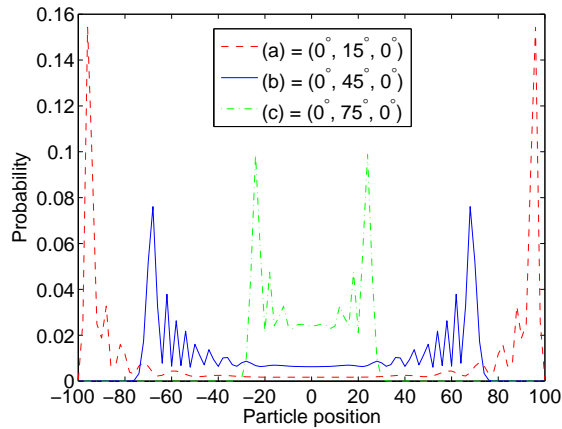, width=8.0cm}
\caption{\label{fig:qw1}Spread of probability distribution for different values of $\theta$ using the quantum coin operator $B_{0, \theta, 0}$. The distribution is wider for (a) = $(0,  \frac{\pi}{12},  0)$ than for (b) = $(0,  \frac{\pi}{4}, 0)$ and (b) = $(0, \frac{5 \pi}{12}, 0)$, showing the decrease in spread with increase in $\theta$. The initial state of the particle $|\Psi_{in}\rangle = (|0\rangle + i |1\rangle) \otimes |j_{0}\rangle$ and the distribution is for 100 steps .}
\end{center}
\end{figure}
\par
The  variance can  be varied by changing the parameter $\theta$, 
\be
\sigma^{2} \approx [ 1-\sin(\theta)]N^{2}.
\ee
The effect of the parameter $\theta$ on the distribution is shown in Fig. \ref{fig:qw1}; the variance decreases with the increase in the value of $\theta$. 
\begin{figure}[h]
\begin{center}
\epsfig{figure=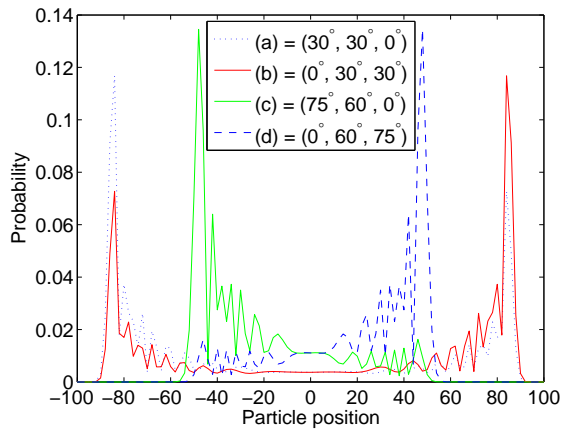, width=8.0cm}
\caption{\label{fig:qw2}Spread of probability distribution for different values of $\xi$, $\theta$, $\zeta$ using quantum coin operator $B_{\xi, \theta, \zeta}$. Biasing the walk using $\xi$ shifts the distribution to the left: (a) = $( \frac{\pi}{6}, \frac{\pi}{6}, 0)$  and (c) = $(\frac{5 \pi}{12}, \frac{\pi}{3}, 0 )$. Biasing the walk using $\zeta$  
shifts it to the right: (b) = $(0, \frac{ \pi}{6}, \frac{\pi}{6})$ and  (d) = $(0, \frac{\pi}{3},  \frac{5 \pi}{12})$. The initial state of the particle $|\Psi_{in}\rangle = (|0\rangle + i |1\rangle) \otimes |j_{0}\rangle$ and the distribution is for 100 steps .}
\end{center}
\end{figure}
The parameters $\xi$ and $\zeta$ introduce asymmetry in the position space probability distribution as shown in Fig. \ref{fig:qw2} and their effect on the variance is negligible \cite{chandra08}.
The simplest coin operation is the Hadamard operator $H = B_{0,45^{\circ},0}$. In this paper, Sec. (\ref{impQW}), we consider the QW with $B_{0,\theta^{\circ},0}$ to redistribute atoms in an optical lattice.
The QW is sensitive to noise \cite{kendon, chandra07, brun} but in Sec. \ref{qwnoise} we consider small amounts of three physically relevant models of noise that can be experimentally induced: a bit-flip channel, a phase-flip channel and an amplitude-damping channel \cite{chandra07}, to act as an enhanced toolbox to control the dynamics of the atoms. 

\section{Phase transition in optical lattice}
\label{pt}
 
A Bose-Einstein condensate (BEC) at low enough temperature is a SF described by a wave function that exhibits long-range coherence \cite{stringari}. When the BEC is transferred to the lattice potential, the atoms move from one lattice site to
the next by tunnel coupling. Using a second-quantized form, the system can be described by the Bose-Hubbard model: \be H = -J
\sum_{\langle j, k \rangle} \hat{b}_{j}^{\dagger}\hat{b}_{k} +
\sum_{j}\epsilon_{j}\hat{n}_{j} + \frac{1}{2}U
\sum_{j}\hat{n}_{j}(\hat{n}_{j}-1), \ee 
\noindent $J$ is the
tunneling term characterized by the hopping amplitude, $\hat{b}_{j}$ 
and $\hat{b}_{j}^{\dagger}$ are the annihilation and
creation operators of atoms, $\epsilon_{j}$ denotes the energy offset 
due to external harmonic confinement of the atoms in the $j$th lattice 
site, $\hat{n}_{j}=\hat{b}_{j}^{\dagger}\hat{b}_{j}$ is the atomic number 
operator counting the number of atoms, and $U$ is the 
repulsive interaction between two atoms in a single lattice site. 
\par
When $J$ dominates the Hamiltonian, the ground-state energy is minimized if the single-particle 
wave functions of all $N$ atoms are spread out over the entire $M$-lattice site.  If $\epsilon_{j} =$ const. 
(homogeneous system) then the many-body ground state is called the SF state and is given by
\be
|\psi_{SF}\rangle_{U=0} \propto \left(\sum_{j=1}^{M} \hat{b}_{j}^{\dagger}\right)^{N}|V\rangle, 
\ee
where $|V\rangle$ is a vacuum state. In this state the probability distribution for the local occupation $n_{j}$ of atoms on a single lattice site is Poissonian. The state is well described by a macroscopic wave function with long-range phase coherence
throughout the lattice. With increase in the ratio $U/J$, the system 
reaches a quantum critical point, the fluctuations in atom number of a 
Poisson distribution become energetically very costly, and the ground state
of the system will instead undergo a quantum phase transition from
the SF state to the MI state, a product of local Fock states of $n$ atoms in each lattice site is given by \cite{jaksch, greiner}, 
\be
|\psi_{MI}\rangle_{J=0} \propto \prod_{j=1}^{M}
(\hat{b}_{j}^{\dagger})^{n}|V\rangle.
\ee

\section{Quantum walk on atoms in 1D Mott Insulator regime}
\label{impQW}

\begin{figure}[h]
\begin{center}
\epsfig{figure=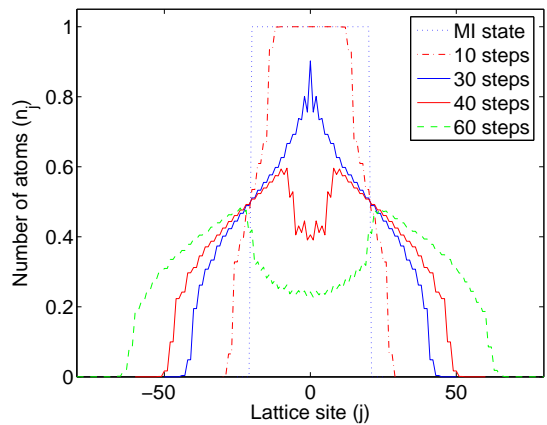, width=8.0cm}
\caption{\label{fig:multiMISF} Density profile of the evolution of 40
atoms starting from MI state in correlation with the position space when subjected to a QW of different numbers of steps with the Hadamard operator $B_{0,45^{\circ},0}$  as the quantum coin. The distribution spreads with increase in the number of steps.}
\end{center}
\end{figure}

Localized atomic wave functions in the MI regime with one atom in each of the $M$ lattice sites are initialized into a symmetric superposition of any of the two internal trappable state, hyperfine levels, $|0\rangle$ and $|1\rangle$ \cite{rfPulse}, 
\be
|\Psi_{MI}\rangle_{J=0} \propto \prod^{\frac{M}{2}}_{j=-\frac{M}{2}}
\left( \frac {|0\rangle + i|1\rangle}{\sqrt{2}}\right)\otimes |j\rangle.
\ee 
The $\mathcal H_{c}$ of each atom is spanned by the two hyperfine levels and $\mathcal H_{p}$ is spanned by the lattice site. The total system is then in the Hilbert space $\mathcal H_{m} = (\Pi_{j}\mathcal H_{c_{j}}) \otimes \mathcal H_{p}$.
The unitary shift operation $S$, Eq. (\ref {eq:condshift}), on the above system will evolve each atom into superposition of the neighbor lattice site, establishing the quantum correlation between the states of the atom and the neighboring lattice site, 
\be
S|\Psi_{MI}\rangle_{J=0} \propto  \prod_{j =
-\frac{M}{2}}^{\frac{M}{2}} \left (
\frac{|0\rangle \otimes \hat{a}|j\rangle +
i|1\rangle \otimes \hat{a}^{\dagger}|j\rangle}{\sqrt 2}\right). 
\ee
To implement the $N$ number of steps, the process of $W = S(B_{\xi,\theta,\zeta} \otimes \mathbbm{1})$ is iterated $N$ times. During the iteration, the atom and position correlations overlap resulting in,
\begin{eqnarray}
\label{eq:manyUMI}
(W)^{N}|\Psi_{MI}\rangle_{J=0}  \propto \prod_{j = -\frac{M}{2}}^{\frac{M}{2}} ( \beta_{j-N}|0\rangle \otimes |j-N\rangle +\nonumber \\
\beta_{j-(N+1)} |0\rangle \otimes |j-(N+1)\rangle +......+\beta_{j+N}|0\rangle \otimes |j+N\rangle \nonumber \\
+ \gamma_{j-N}|1\rangle \otimes |j-N\rangle + ......+ \gamma_{j+N})|1\rangle \otimes  |j+N\rangle ).
\end{eqnarray}
\noindent This can be written as
\begin{eqnarray}
\label{eq:manyUMIa}
(W)^{N}|\Psi_{MI}\rangle_{J=0}  \propto  
\prod_{j = -\frac{M}{2}}^{\frac{M}{2}}\left( \sum\limits_{x=j-N}^{j+N} [ \beta_x |0\rangle + \gamma_x |1\rangle ] \otimes |x\rangle \right ),  
\end{eqnarray}
\noindent $\beta_x$ and $\gamma_x$ are the probability amplitudes of state $|0\rangle$ and $|1\rangle$ at lattice site $x$, which range from $(j-N)$ to $(j+N)$. For a QW of $N$ steps on a particle initially at position $j=0$ using $B_{0, \theta, 0}$ as the quantum coin, the probability distribution spread over the interval $(-N \cos(\theta), N \cos(\theta))$ in position space and shrink quickly outside this region \cite{ashwin, chandra08}. Therefore, after $N$ steps the density profile of $M$ atoms initially between $\pm \frac{M}{2}$ will correlate with the position space and spread over the lattice site $\pm (\frac{M}{2} + N \cos(\theta))$.
 Figure \ref{fig:multiMISF} is the redistribution of atomic density of 40 atoms initially in the MI state when subjected to a QW of different numbers of steps with Hadamard operator $H = B_{0,45^{\circ},0}$ as the quantum coin.
\par
The redistributed atoms would be in either of their internal states and this can be retained to study the phase transition of the two-state bosonic atoms in an optical lattice \cite{altman} or all atoms can be transferred to one of the internal states. A technique based on adiabatic passage using crafted laser pulses can be used for a nearly complete transfer of population between two states \cite{Vitanov01}, and a popular example of one such technique is stimulated Raman adiabatic passage (STIRAP) \cite {stirap}. 
\par
Along with the correlation between the atoms and the position space, Eq. (\ref{eq:manyUMI}) also reveals the overlap of the probability amplitude of different atoms in the position space. That is, during each step of the QW, the amplitude of the states of each atom overlaps with the amplitude of the states of the atoms in the neighboring lattice site.  After the iteration of the $N$ steps $ \geq \frac{M/2}{\cos(\theta)}$, the overlap of all atoms could be seen at the central region of the lattice. 
\par
When the number of steps of the QW is equal to the number of lattice site ($N=M$), the overlap of the fraction of amplitude of all $M$ atoms exists within the lattice site $\pm\frac{M\cos(\theta)}{2}$ making the region identical to the SF state. Beyond $\pm\frac{M\cos(\theta)}{2}$ the number of atoms in correlation with that position space decreases and hence the number of overlapping atoms also decrease.
To make all $M$ atoms to spread between lattice site $\pm\frac{M}{2}$ using the usual method of lowering the optical potential depth following the CRW protocol takes $M^{2}$ steps. Therefore, we can conclude that using QW a long range correlation can be induced quadratically faster than by the usual technique.
\begin{figure}[h]
\begin{center}
\epsfig{figure=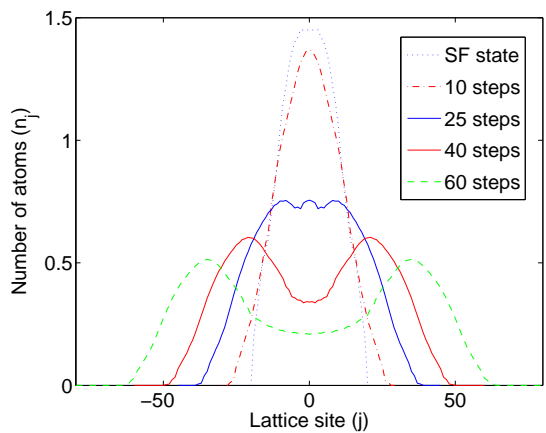, width=8.0cm}
\caption{\label{fig:multiSFMI} Density profile of the evolution of 40
atoms starting from SF state in correlation with the position space when subjected to a QW of different numbers of steps with the Hadamard operator $B_{0,45^{\circ},0}$  as the quantum coin. The distribution spread with increase in the number of steps and is almost uniform between the lattice site $\pm 20$ when $N = 25$. At this stage the optical potential depth can be increased to cancel the correlation and obtain the MI state.}
\end{center}
\end{figure}

\par Similarly,  Fig. \ref{fig:multiSFMI} is the redistribution of atomic density of 40 atoms initially in the SF states when subjected to the QWs of different number of steps using Hadamard operator $B_{0,45^{\circ},0}$ as the quantum coin. To reach MI state the distribution should be uniform over the lattice site without correlation. A distribution that is approximately uniform within the region $\pm \frac{M}{2}$ can be obtained using the QW.  In Fig. (\ref{fig:multiSFMI}) when $N=25$ the distribution is almost uniform between the lattice site $\pm 20$ retaining the correlation with the position space.  The uniformity of the distribution can be improved by using the tunnelling of atoms between the lattice. Once the distribution is uniform, the optical potential depth can be increased to cancel the correlation and obtain the MI state. 
The uniformity of the distribution can also be improved by introducing a noise channel, as we show later. 
\par
\begin{figure}[h]
\begin{center}
\epsfig{figure=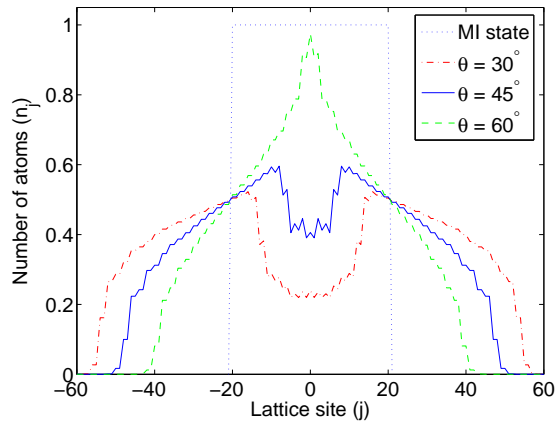, width=8.0cm} 
\caption{\label{fig:multi1}Distribution of atoms initially in MI state when subjected to a QW ($N=M=40)$ using different values for $\theta$ in the coin operator $B_{0, \theta, 0}$. The spread is wider for $\theta = 30^{\circ}$ and decreases with increasing $\theta$. }
\end{center}
\end{figure}
\begin{figure}[h]
\begin{center}
\epsfig{figure=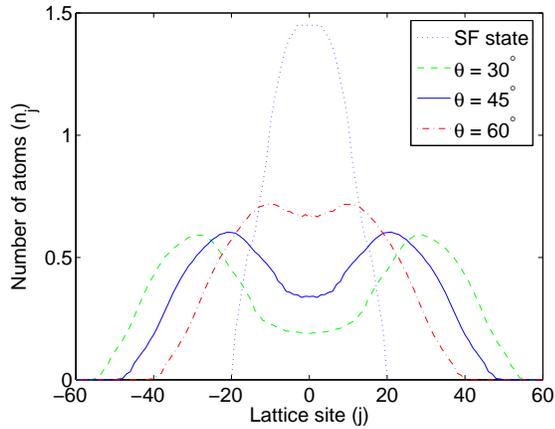, width=8.0cm}
\caption{\label{fig:multi2}Distribution of atoms initially in SF state when subjected to a QW ($N=M=40)$ using different values for $\theta$ in the coin operator $B_{0, \theta, 0}$. The distribution is almost uniform for $\theta=60^{\circ}$ between the lattice site $\pm 20$. At this value the optical potential depth can be increased to cancel the correlation and obtain the MI state.}
\end{center}
\end{figure}

The variance and the probability distribution can be controlled using the parameters $\theta$, $\xi$ and $\zeta$ in $B_{\xi, \theta, \zeta}$. Fig. (\ref{fig:multi1}, \ref{fig:multi2}) shows the density distribution obtained by implementing the QW on atoms in MI and SF state $(N=M=40)$ with different values of $\theta$ in the coin operator $B_{0,\theta,0}$.  The spread is wider for $\theta=30^{\circ}$ and decreases with increase in $\theta$. In Fig. (\ref{fig:multi2}) the distribution is almost uniform for $\theta=60^{\circ}$ between the lattice site $\pm 20$. At this value the optical potential depth can be increased to cancel the correlation and obtain the MI state.
\par
\par
The Hamiltonian of the system in general can be described by the 1D Bose-Hubbard model for two-state atoms, 
\begin{eqnarray}
\label{eq:manyH}
H = -J_{\uparrow}\sum_{\langle j, k \rangle} \hat{b}_{j \uparrow}^{\dagger}\hat{b}_{k \uparrow} -J_{\downarrow}\sum_{\langle j, k \rangle} \hat{b}_{j \downarrow}^{\dagger}\hat{b}_{k \downarrow} +  \sum_{j, \alpha = \uparrow , \downarrow}\epsilon_{j, \alpha}\hat{n}_{j, \alpha} + \nonumber \\
U \sum_{j}(\hat{n}_{j \uparrow} -\frac{1}{2})(\hat{n}_{j \downarrow}-\frac{1}{2})+  \frac{1}{2}\sum_{j, \alpha = \uparrow , \downarrow}V_{\alpha} \hat{n}_{j \alpha}(\hat{n}_{j \alpha}-1) \nonumber \\
+ \sum_{j}\left ( d_{L} \hat{b}_{(j-1)\uparrow}^{\dagger}\hat{b}_{j\uparrow} + d_{R} \hat{b}_{(j+1)\downarrow}^{\dagger}\hat{b}_{j\downarrow } \right),
\end{eqnarray}
\noindent
here $\uparrow$ and $\downarrow$ represent the terms for atoms in state $|0\rangle$ and state $|1\rangle$ respectively. $U$ is the interaction between atoms in state $|0\rangle$ and $|1\rangle$; $V_{\uparrow (\downarrow)}$ is the interaction between atoms in same state. $d_L$ and $d_R$ are the left and right displacement terms induced during each step of the QW. The Hamiltonian is evolved with coin toss operation in a regular interval of time $t$, the time required to move the atom to the neighboring site. The optical lattice can be dynamically manipulated to evolve atoms in a superposition of lattice site without giving time for the atom-atom interaction, and the optical potential depth of the system can be configured just above the level where there is no direct tunneling. Then the dynamics of atoms in lattice will be dominated by the last term of the Hamiltonian in Eq. (\ref{eq:manyH}) ignoring the  atom-atom interaction. Therefore scaling up the scheme of using the QW on systems with large number of atoms in each lattice site or to infinitely large numbers of lattice site is also straight forward; whereas, the atom-atom interaction play a prominent role during the usual method of varying the potential depth.

\section{Quantum walk with a noisy channel as a toolbox}
\label{qwnoise}

Along with the coin parameters $\theta$, $\xi$ and $\zeta$, a small amount of engineered noise can be used to control the redistribution of atom in the optical lattice. To demonstrate the effect of the QW with a noisy channel on atoms in optical lattice we consider a bit-flip channel, a phase-flip channel and an amplitude-damping channel.  The bit-flip channel flips the state of the particle from $|0\rangle$ ($|1\rangle$) to $|1\rangle$ ($|0\rangle$) (Pauli $X$ operation) and a phase-flip channel flips the phase of $|1\rangle$ to $-|1\rangle$ (Pauli $Z$ operation). We use the notation $p$ for the noise level where, $0\le p \le 1$. Therefore, the bit-  (phase-)flip channel flips the state (phase) with probability $(1-p)$ during each step of the QW.  
An amplitude-damping channel leaves state $|0\rangle$ unchanged but reduces the amplitude of state $|1\rangle$ with probability $(1-p)$ resulting in an asymmetric distribution. The maximum decoherence effect using 
bit- and phase-flip channel is at $p=0.5$ due to symmetries induced by these two channels during the QW; whereas, the amplitude-damping channel does not obey any symmetry and hence the maximum effect is for $p=1$ \cite{chandra07}. Our numerical implementation of these channels evolves the density matrix employing the Kraus operator representation. 
\par
\begin{figure}
\begin{center}
\epsfig{figure=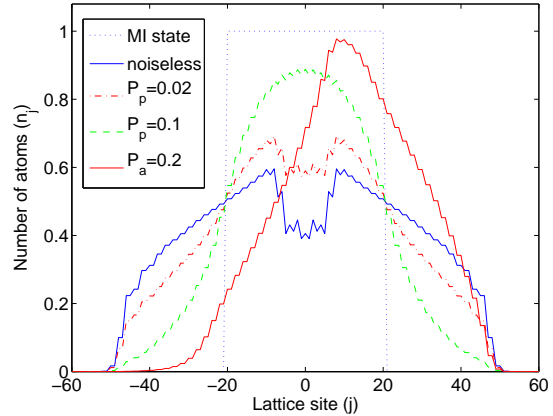, width=8.0cm}
\caption{\label{fig:MISFPAdamp} Atoms in MI state after implementing the QW ($N = M = 40, \theta=45^{\circ})$ with noise channel. With increased phase damping from $p_{p} = 0.02$ to $p_{p} = 0.1$ the distribution ($n_{j}$) at the central region gets closure to Gaussian. Amplitude damping of state $|1\rangle$ introduces asymmetry to the distribution, $p_{a} = 0.2$.}
\end{center}
\end{figure}
\begin{figure}
\begin{center}
\epsfig{figure=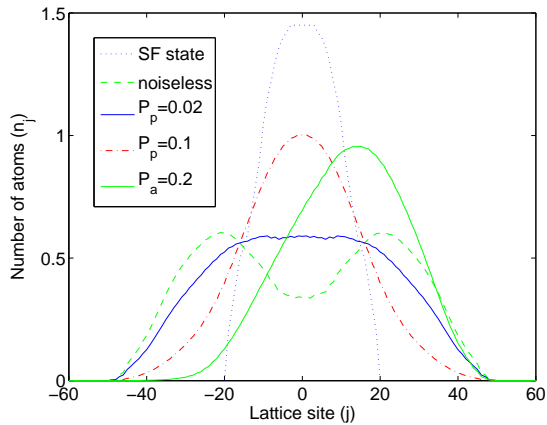, width=8.0cm}
\caption{\label{fig:multi5}
Atoms in SF state after implementing a QW $(N = M = 40, \theta=45^{\circ})$. With noiseless QW, the  atoms spread, getting close to a uniform distribution. For phase damping $p_{p} = 0.02$ the distribution is uniform between $\pm 20$ and gets closer to Gaussian at $p_{p} = 0.1$. Amplitude damping of state $|1\rangle$ introduces asymmetry to the distribution, $p_{a} = 0.2$.}
\end{center}
\end{figure}

Figure \ref{fig:MISFPAdamp} is a comparison of the distribution obtained without noise channels to the distributions with phase-flip ($p = 0.02$ and $p=0.1$) and amplitude-damping channel ($p=0.2$) on atoms initially in the MI state. 
Similarly, the redistribution of atoms in SF state when subjected to QW without and with noise channels is shown in Fig. \ref{fig:multi5}. With the addition of phase flip noise 
of $p=0.02$, a uniform distribution is obtained. Then the optical potential depth can be increased to cancel the correlation and obtain MI state. 
\par
The effect of bit-flip on the distribution is close to the one obtained using phase-flip channel and hence the redistribution in the Figs. \ref{fig:MISFPAdamp} and \ref{fig:multi5} are shown only for phase-flip and amplitude-damping channels. Increasing noise level affects the variance of the QW \cite{chandra07}, which proportionally affect the atom-position correlation and the atom-atom overlap region. Therefore, it is important to restrict the noise level to a very low value ($p \lessapprox 0.1$) to use noisy channel as a tool.
\par
Some of the distribution presented in Secs. \ref{impQW} and \ref{qwnoise} using QWs are visibly similar to some of the distribution presented using different technique; in \cite{rodriguez} using time evolution density matrix renormalization group (t-DMRG) and in \cite{sengupta, marcos} using quantum Monte Carlo simulation. These technique present the change in density profile distribution with time, whereas in this paper we have discussed the change with number of steps of the QW. 

\section{implementation}
\label{impl}

Optical lattices ranging from a simple periodic, square and cubic, to a more exotic ones, such as hexagonal  and Kagome lattices using superlattice technique \cite{duan} have been created to trap and manipulate cold atoms. Manipulation of cold atoms in time-varying optical lattice has also been reported \cite{boyer}. This has provided flexibility in designing and studying quantum phases and quantum phase transitions in coherent and strongly correlated ultracold atoms. 

In Ref. \cite{mandel}, controlled coherent transportation of rubidium atoms in the spin-dependent optical lattice has been experimentally demonstrated. Rubidium atoms in the MI state is evolved into superposition of two internal state and the two states are transported in opposite directions.  Delocalization of atoms over the seven lattice sites is demonstrated. This scheme can be adopted to implement the QW and to redistribute the density profile by introducing a $rf$-pulse \cite{rfPulse} after each separation. An additional $rf$-pulse can be engineered to act as noise channels. 
\par
In an implementation scheme proposed in Ref. \cite{chandra06}, a {\em  stimulated Raman kick}, two selected levels of the atom are coupled to the two modes of counterpropagating laser beams to impart a translation of atoms in the position space. To realize transition from the SF to the MI phase using this scheme, atoms are first redistributed by implementing the QW in a long-Rayleigh-range optical trap \cite{rayleigh}  and the optical lattice is switched on at the end to confine atoms and cancel the correlation. For transition from MI to SF, the atoms initially in the optical lattice are transferred to long-Rayleigh-range dipole trap and a QW is implemented. 

\section{conclusion}
\label{conc}

In this paper, we have shown the use of the QW to study the dynamics of atoms in an optical lattice and expedite the process of quantum phase transition. We have also used the QW with experimentally realizable noisy channels to show the additional control one can have over the evolution and atomic density redistribution.  Theoretically, the evolution of the density profile with a QW can be used in place of quantum Monte Carlo simulation to study the correlation and redistribution of atoms in optical lattices. We expect the QW to play a wider role in simulating and expediting the dynamics in various physical systems.

\bc
{\bf Acknowledgement}
\ec
We thank Gerardo Ortiz for helpful discussion. This work was funded by CIFAR, ARO, and MITACS. CMC would like to thank Mike and Ophelia Lezaridis  for the financial support at IQC.



\begin{thebibliography}{99}
\bibitem{rfPulse} A $\pi/2$ rf pulse can evolve the atoms into equal superposition of the two internal trappable states.
\bibitem{rayleigh} Distance at which the diameter of the spot size increases by a factor of  $\sqrt{2}$.
\end{thebibliography}
\end{document}